\documentclass[english,prl,twocolumn,aps,amssymb,nofootinbib]{revtex4-1}
\usepackage[T1]{fontenc}
\usepackage{amsmath}
\usepackage{amssymb}
\usepackage{graphicx}
\usepackage{braket}
\usepackage{babel}
\usepackage{color}

\usepackage{braket}

\usepackage[T1]{fontenc}
\usepackage[latin9]{inputenc}
\usepackage{textcomp}
\usepackage{color}

\usepackage{babel}
\usepackage[symbol]{footmisc}

\makeatother

\begin{document}

\title{Single electron-spin-resonance detection by microwave photon counting}

\author{Z. Wang$^{  1,2,*}$, L. Balembois$^{ 1, *}$, M. Rancic$^1$, E. Billaud$^1$, M. Le Dantec$^1$, A. Ferrier$^3$, P. Goldner$^3$, S. Bertaina$^4$, T. Chaneliere$^5$, D. Esteve$^1$, D. Vion$^1$, P. Bertet$^1$, E. Flurin$^{1,\dagger}$}

\footnotetext[1]{these authors contributed equally}
\footnotetext[2]{corresponding author: emmanuel.flurin@cea.fr}

\affiliation{
\\ \
\\
$^1$Quantronics group, Universit\'e Paris-Saclay, CEA, CNRS, SPEC, 91191 Gif-sur-Yvette Cedex, France\\
$^2$D\'epartement de Physique et Institut Quantique, Universit\'e de Sherbrooke, Sherbrooke, Qu\'ebec, Canada\\
$^3$Chimie ParisTech, PSL University, CNRS, Institut de Recherche de Chimie Paris, 75005 Paris, France\\
$^4$CNRS,  Aix-Marseille  Universit\'e,  IM2NP  (UMR  7334),  Institut  Mat\'eriaux Micro\'electronique  et  Nanosciences de Provence,  Marseille, France \\
$^5$Univ. Grenoble Alpes, CNRS, Grenoble INP, Institut N\'eel, 38000 Grenoble, France}

\date{\today}

\maketitle

\textbf{Electron spin resonance (ESR) spectroscopy is the method of choice for characterizing paramagnetic impurities, with applications ranging from chemistry to quantum computing~\cite{schweiger_principles_2001}, but it gives access only to ensemble-averaged quantities due to its limited signal-to-noise ratio. Single-electron-spin sensitivity has however been reached using spin-dependent photoluminescence~\cite{wrachtrup_optical_1993,gruber_scanning_1997,raha_optical_2020}, transport measurements~\cite{elzerman_single-shot_2004,vincent_electronic_2012,pla_single-atom_2012,thiele_electrically_2014}, and scanning-probe techniques~\cite{rugar_single_2004,baumann_electron_2015,grinolds_subnanometre_2014}. These methods are system-specific or sensitive only in a small detection volume, so that practical single spin detection remains an open challenge. Here, we demonstrate single electron magnetic resonance by spin fluorescence detection~\cite{albertinale_detecting_2021}, using a microwave photon counter at cryogenic temperatures~\cite{lescanne_irreversible_2020}. We detect individual paramagnetic erbium ions in a scheelite crystal coupled to a high-quality factor planar superconducting resonator to enhance their radiative decay rate, with a signal-to-noise ratio of $1.9$ in one second integration time. The fluorescence signal shows anti-bunching, proving that it comes from individual emitters. Coherence times up to $3$\,ms are measured, limited by the spin radiative lifetime. The method has the potential to apply to arbitrary paramagnetic species with long enough non-radiative relaxation time, and allows single-spin detection in a volume as large as the resonator magnetic mode volume ($\sim 10 \mu \mathrm{m}^3$ in the present experiment), orders of magnitude larger than other single-spin detection techniques. As such, it may find applications in magnetic resonance and quantum computing.
}

In ESR spectroscopy, the linewidth of an ensemble of paramagnetic centers is usually dominated by the frequency shifts that each center undergoes under the action of its local environment. This inhomogeneous broadening can reach large values (up to several GHz) and imposes a limitation to the achievable spectral resolution~\cite{schweiger_principles_2001}. One radical way to overcome the inhomogeneous broadening is to perform ESR spectroscopy on individual paramagnetic centers, thus gaining several orders of magnitude in spectral resolution since single spin linewidths are typically in the kHz-MHz range~\cite{gruber_scanning_1997,pla_single-atom_2012,muhonen_storing_2014}. Besides the interest for magnetic resonance spectroscopy, single spin addressing is also a necessity for most spin-based quantum computing applications.  

Practical single-electron-spin-resonance should enable the detection and spectroscopy of a wide range of paramagnetic centers buried in an insulating matrix, with a sufficiently large detection volume and signal-to-noise ratio. So far, none of the approaches that achieve single-spin detection satisfy all of these requirements. Optically Detected Magnetic Resonance (ODMR) can detect individual paramagnetic centers only when suitable energy levels and cycling optical transitions are present~\cite{wrachtrup_optical_1993,gruber_scanning_1997,raha_optical_2020}. ODMR-detected individual NV centers can be used to measure the spectrum of neighboring single electron spins in ambient conditions~\cite{grinolds_subnanometre_2014,shi_single-protein_2015,shi_single-dna_2018}, but the detection volume is limited to $\sim 10^3 - 10^4 \mathrm{nm}^3$ by the $1/r^3$ dependence of the dipolar interaction, which makes the detection of spins far outside of the diamond host challenging. Spin-dependent transport can detect individual spins when a spin-to-charge conversion pathway is present~\cite{elzerman_single-shot_2004,vincent_electronic_2012,pla_single-atom_2012,thiele_electrically_2014,baumann_electron_2015}, but this is lacking in most paramagnetic centers. Single electron-spin imaging was also achieved using Magnetic Resonance Force Microscopy~\cite{rugar_single_2004}, but spectroscopy has not yet been demonstrated with this platform.
\begin{figure}[!tbh]
\includegraphics[width=0.45\textwidth]{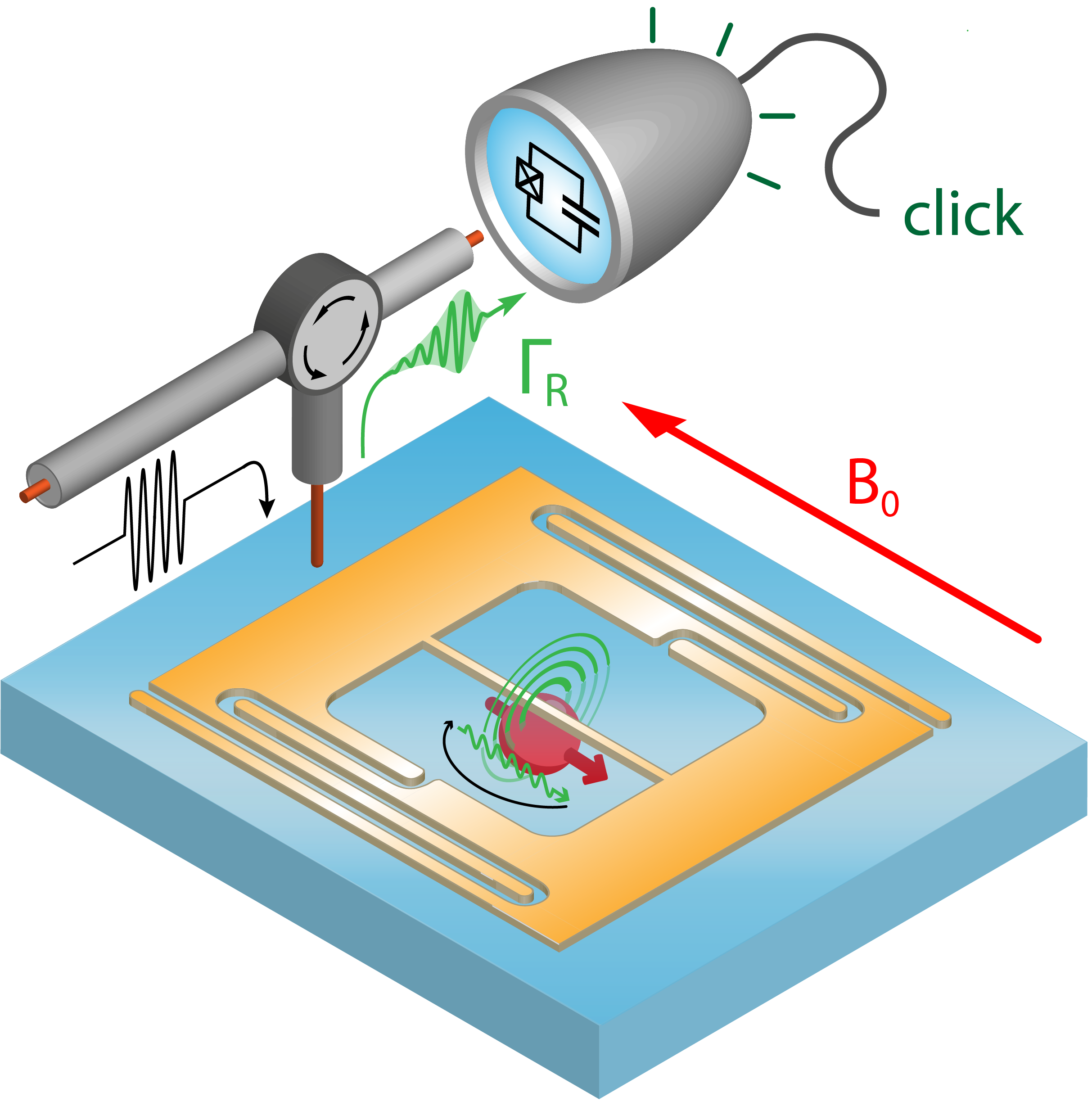}
\caption{
\textbf{Principle of single spin spectroscopy by microwave photon counting}. An individual electron spin  (red arrow) embedded in a crystal is excited by a microwave pulse (in black); it then relaxes back to its ground state by emitting a microwave photon (green arrow), which is routed via a circulator towards a microwave photon counter based on a superconducting transmon qubit. To enhance its radiative rate $\Gamma_R$, the spin is coupled magnetically to the mode of a high-quality-factor superconducting planar microwave LC resonator (in orange). The spin frequency is tuned to the resonator by application of a magnetic field $B_0$ parallel to the resonator plane.  
}
\label{fig1}
\end{figure}

\begin{figure*}[!tbh]
\includegraphics[width=1\textwidth]{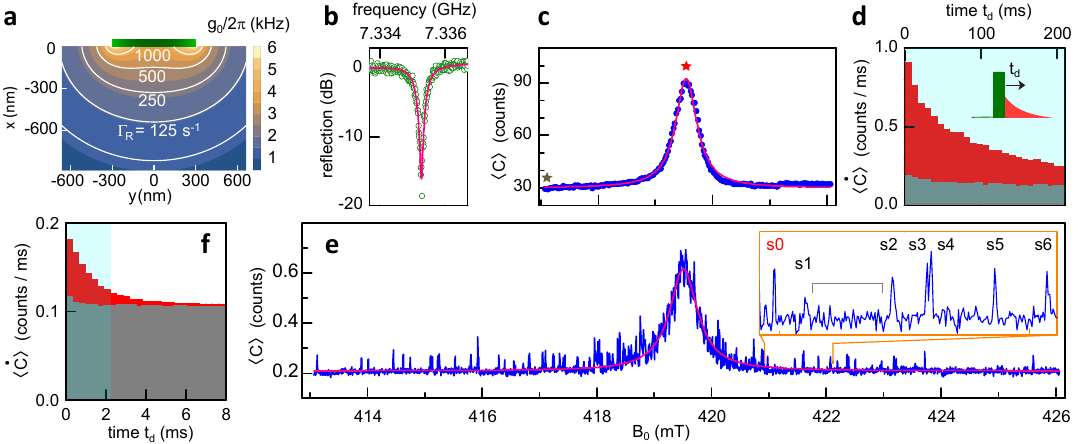}
\caption{
\textbf{Spin spectroscopy.} 
\textbf{(a)} Simulation of the spin-resonator coupling constant $g_0(x,y)$ and relaxation rate $\Gamma_R(x,y)$ as a function of the spin position $(x,y)$ with respect to the wire (shown as a green rectangle). \textbf{(b)} Magnitude of resonator reflection (green dots) as a function of probe frequency at single-photon level input power. A fit yields the total resonator linewidth $\kappa/2\pi=470$\,kHz, with a coupling rate $\kappa_c = 1.7 \cdot 10^6\,\mathrm{s}^{-1}$ and an internal loss rate $\kappa_i= 1.3 \cdot 10^6\,\mathrm{s}^{-1}$ \textbf{(c-d)} Microwave fluorescence spectroscopy (c) at high excitation power ($\sim - 97$\,dBm at sample input) and typical fluorescence signal (d). At each magnetic field $B_0$, the average number of counts $\langle C \rangle$ is integrated over a $\sim200$\,ms duration following the excitation pulse (light blue window in panel d). Blue open circles are data, red line is a Lorentzian fit with FWHM $0.45$\,mT. Note that the angle $\theta$ varies linearly between $-0.006^{\circ}$ and $0.006^{\circ}$ over the scan. Fluorescence histograms are shown at the center (red) and tail (grey) of the spin ensemble line (see stars in panel c).  \textbf{(e)} Spin spectroscopy at low power ($\sim - 107$\,dBm at sample input), with an integration window of $2$\,ms. Blue line is measured data, red line is a Lorentzian fit. The inset shows an expanded view of 7 peaks (labelled $s0$ to $s6$). Note that the angle $\theta$ varies linearly between $-0.016^{\circ}$ and $0.016^{\circ}$ over the scan. \textbf{(f)} Fluorescence histograms of spin $s0$ (red) and background (grey) averaged over the range of $B_0$ shown in the inset of panel e. The light blue window is the integration window for the data in e). 
}
\label{fig2}
\end{figure*}

Here, we perform single electron spin resonance spectroscopy by transposing fluorescence detection, a well-established method to detect individual emitters in the optical domain at room-temperature, to microwave frequencies and millikelvin temperatures. In optical fluorescence, an emitter is excited by a short light pulse, and detected by counting the emitted photons during the radiative relaxation~\cite{orrit_single_1990,gruber_scanning_1997}. Similarly, we excite a spin by a short microwave pulse, and detect it by counting the microwave photons it emits when returning to its ground state. Spin relaxation by spontaneous emission of microwave photons is exceedingly slow in free space; we thus enhance its rate $\Gamma_R$ resonantly by coupling the spin to a high-quality-factor superconducting microwave resonator of frequency $\omega_0$~\cite{bienfait_controlling_2016}, and we detect the fluorescence photon with a single-microwave-photon detector (SMPD) based on a superconducting qubit (see Fig.~\ref{fig1} for a schematic description). The maximum signal-to-noise ratio (SNR) reached by this method with a one second integration time scales as $\sim \eta \Gamma_R / \sqrt{\alpha + \eta (1-\eta)\Gamma_R}$, where $0\leq \eta \leq 1$ is the average number of counts generated by the radiative decay of one spin, and $\alpha$ the SMPD dark count rate (see Methods). It is noteworthy that this SNR is only limited by technical imperfections and has no upper bound for an ideal experiment where $\alpha=0$ and $\eta=1$, in contrast with earlier proposals and experiments of circuit-QED-enhanced magnetic resonance where the SNR is ultimately limited by vacuum microwave fluctuations ~\cite{kubo_electron_2012,bienfait_reaching_2016,haikka_proposal_2017,eichler_electron_2017,budoyo_electron_2018,budoyo_electron_2020,ranjan_electron_2020}. In a recent experiment demonstrating the detection of $\sim 10^4 $ impurity spins by microwave fluorescence, this single-spin SNR was $\sim  5 \cdot 10^{-4}$~\cite{albertinale_detecting_2021}, thus insufficient for single-spin detection. Here, we reach a single-spin SNR of $\sim 1$ by improving the resonator design, the SMPD performance, and by using spins with a larger gyromagnetic ratio. Our method could be applicable to a broad class of paramagnetic impurities and offers a detection volume that can be large ($\sim 10 \,\mu \mathrm{m}^3 $ in the present experiment). It is therefore promising for operational single electron spin resonance at cryogenic temperatures. 

We demonstrate this method with rare-earth ions in a crystal, specifically $\mathrm{Er}^{3+}$ ions in a scheelite crystal of $\mathrm{CaWO}_4$, which has tetragonal symmetry around its $c$-axis. The  crystal used in the experiment was grown undoped, but has a residual erbium concentration $3.1 \pm 0.2 $\ ppb (see Methods), which corresponds to a $\sim 300\, \mathrm{nm}$ average distance between neighboring $\mathrm{Er}^{3+}$ ions. At low temperatures, only the ground state Kramers doublet of $\mathrm{Er}^{3+}:\mathrm{CaWO}_4$ is populated; it behaves as an effective spin $S=1/2$ with frequency $\omega_s  = \mathbf{\gamma} \cdot \mathbf{B_0}$, where $\mathbf{B_0}$ is the applied magnetic field and $\mathbf{\gamma}$ the ion gyromagnetic tensor. The ensemble-averaged gyromagnetic tensor $\mathbf{\gamma}_0$ determines the center of the ensemble resonance line $\omega_{s0}$; it is diagonal in the $(a,b,c)$ tetragonal frame, with elements $\gamma_a = \gamma_b \equiv \gamma_\perp = 2\pi \times 117.3$\,GHz/T, and $\gamma_{c} \equiv \gamma_{||} = 2\pi \times 17.45$\,GHz/T \cite{antipin_a._paramagnetic_1968}. Due to inhomogeneous broadening however, each individual ion has a gyromagnetic tensor $\mathbf{\gamma} = \mathbf{\gamma}_0 + \mathbf{\delta \gamma}$ (with $|\mathbf{\delta \gamma}| \ll |\mathbf{\gamma}_0|$) that slightly deviates from $\mathbf{\gamma}_0$~\cite{mims_broadening_1966}. 

\begin{figure}[tbh!]
\centerline{\includegraphics[width=0.5\textwidth]{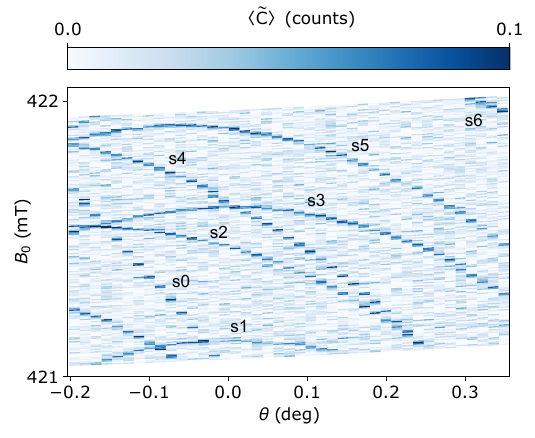}}
\caption{
\textbf{Single-spin-resolved rotation pattern} Average number of excess count $\langle \widetilde{C} \rangle$ as a function of the magnetic field amplitude $B_0$ and its angle $\theta$ with respect to the projection of the crystal $c$ axis on the sample surface. The range is the same as in the inset of Fig.~\ref{fig2}, and the same labeling of the spin lines is used. The data acquisition time was approximately one week.
}
\label{fig3}
\end{figure}

The planar resonator is patterned on top of the crystal, out of a superconducting niobium thin-film. The heart of the device is a $600\,\mathrm{nm}$-wide, $100 \mu \mathrm{m}$-long wire, which acts as a lumped inductance, shunted by a finger capacitor (see Fig.~\ref{fig1} and Methods) that sets the resonance frequency $\omega_0/2\pi = 7.335$\,GHz. The wire ($z$ direction) is oriented approximately along the crystal $c$-axis, and the magnetic field $B_0$ is applied along the sample surface ($z - y$ plane), at a small adjustable angle $\theta$ with respect to $z$ (see Methods). The resonator is coupled to a transmission line for exciting the spins and collecting their fluorescence, at a rate $\kappa_c$, whereas the total resonator damping rate $\kappa = \kappa_c + \kappa_i$ also includes internal losses $\kappa_i$. A circulator routes the excitation pulses from the input line towards the sample, and the reflected pulses together with the subsequent spin fluorescence signal towards the input of a transmon-qubit-based SMPD. This detector is similar to the ones described in~\cite{lescanne_irreversible_2020,albertinale_detecting_2021}, but has a much lower dark count rate $\alpha = 10^2\, \mathrm{s}^{-1}$ (see Methods).

By coupling to the resonator, a spin at frequency $\omega_s$ and position $\mathbf{r}$ demonstrates a Purcell-enhanced radiative relaxation rate $\Gamma_R = \kappa g_0^2 / (\delta^2 + \kappa^2/4)$ that depends on its detuning to the resonator $\delta \equiv \omega_s - \omega_0$ and on the magnetic field vacuum fluctuations $\mathbf{\delta B_1}(\mathbf{r})$ through the spin-resonator coupling strength $g_0 (\mathbf{r}) = \langle \downarrow | \mathbf{S}| \uparrow \rangle \cdot \mathbf{\gamma} \cdot \mathbf{\delta B_1}(\mathbf{r})$~\cite{bienfait_controlling_2016}. To a good approximation, $\mathbf{\delta B_1}(\mathbf{r})$ does not depend on the position $z$ along the wire and is orthogonal to the latter, so that the coupling strength can be re-written as $g_0 (x,y) = (1/2) \gamma_\perp |\mathbf{\delta B_1}(x,y)|$. The $g_0 (x,y)$ map is shown in Fig.~\ref{fig2}a for our resonator design, and shows that $g_0/2\pi$ is larger than $3$\,kHz, and thus $\Gamma_R$ is larger than $\sim 500 \mathrm{s}^{-1}$, for spins located below the wire at a depth smaller than $\sim 150 \,\mathrm{nm}$, corresponding to a volume of $\sim 10\,\mu \mathrm{m}^3$. This implies that $\Gamma_R / \sqrt{\alpha} \geq 50/\sqrt{\mathrm{Hz}}$, and therefore suggests that single-spin sensitivity may be reached over this whole volume.

The properties of the fluorescence signal, which is the sum of the contributions of all the spins excited by the pulse, strongly depend on the excitation power. We first record the spectrum of the $\mathrm{Er}^{3+}:\mathrm{CaWO}_4$ resonance with a high input power ($\sim -97$\,dBm), thus exciting many weakly coupled ions that have low $\Gamma_R$. The average count rate as a function of time following the pulse shows an excess compared to the dark count level (see Fig.~\ref{fig2}d) and decays non-exponentially over a time scale of $\sim 100$\,ms. We plot the average number of counts integrated over $200$\,ms $\langle C \rangle$ as a function of magnetic field $B_0$ applied along the $z$ direction in Fig.~\ref{fig2}c. A smooth, approximately Lorentzian, peak is observed at $B_0 = 419.5$\,mT, close to $\omega_0 / \gamma_{||}$, proving it is the $\mathrm{Er}^{3+}$ spin resonance. Its inhomogeneous Full-Width-Half-Maximum linewidth $0.5$\,mT corresponds to a $\sim 8$\,MHz-wide distribution.

We then record the line with $\sim 20$\,dB lower excitation power while simultaneously reducing the integration time to $2$\,ms, thus exciting and detecting only the most strongly coupled and fastest relaxing spins. The integrated count $\langle C \rangle (B_0)$ now shows qualitatively different behavior and appears as a sum of narrow, unevenly distributed peaks, with typical amplitude $\sim 0.1$ excess count over the noise floor (see Fig.\ref{fig2}e). The fluorescence curve when tuned to one of these peaks shows an exponential decay (see Fig.~\ref{fig2}f), with a time constant of $\sim 2$\,ms. These features suggest that each peak corresponds to the microwave fluorescence signal originating from a single $\mathrm{Er}^{3+}$ ion spin; analogous to the optical fluorescence spectrum of a collection of individual solid-state emitters~\cite{orrit_single_1990,kindem_control_2020,dibos_atomic_2018}. Note that while we observe a large fluorescence signal at the centre of the inhomogeneous absorption line, some individual peaks are still found far from the centre; a common observation in low-density spectra of optical emitters, and a natural consequence of the random nature of inhomogeneous broadening. This is also possibly supplemented in our particular device by the strain imparted by the thermal contractions of the metallic wire on the substrate just below~\cite{pla_strain-induced_2018,ranjan_spatially-resolved_2021}.

\begin{figure}[tbh!]
\centerline{\includegraphics[width=0.48\textwidth]{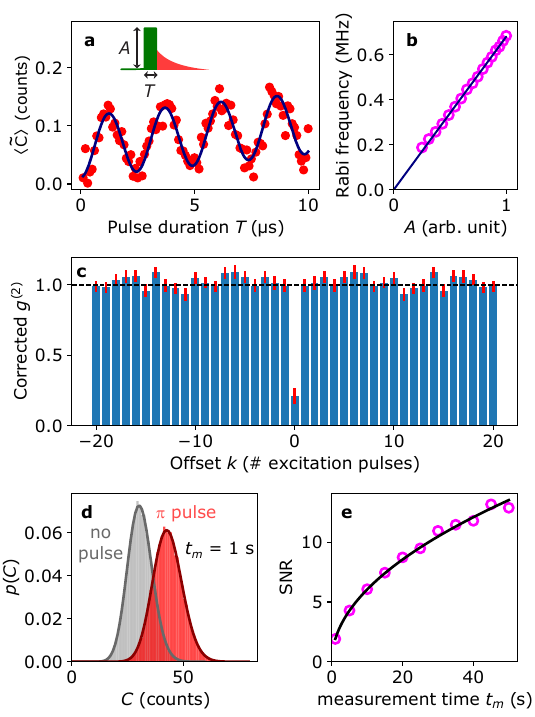}}
\caption{
\textbf{Characterization of spin $s0$.}
\textbf{(a)} Rabi oscillation: measured average excess count $\langle \widetilde{C}\rangle$ (red dots) as a function of excitation pulse duration T (see inset), and corresponding fit (solid line) by a sine function with linearly increasing offset. \textbf{(b)} Extracted Rabi frequency (magenta dots) as a function of excitation pulse amplitude $A$, and corresponding linear fit through origin (solid line). \textbf{(c)} Background-corrected auto-correlation function $g^{(2)}$ (blue columns) and corresponding $\pm1$-standard deviation error bars (red) measured as a function of the offset $k$ between excitation pulses.
\textbf{(d)} Measured probability distribution $p(C)$ of the total count $C$ integrated over the first $2$\,ms of 7.5 ms-long sequences, either with no excitation pulse applied (grey) or with a $\pi$ excitation pulse (red). Sequences are repeated and counts are summed during a measurement time $t_m = 1$\,s. Solid lines are Poissonian fits, yielding the spin signal $C_{spin} = 12.4$ (difference between the mean values of the two distributions) and the standard deviations $\delta C_0 = 5.5$ and $\delta C_{\pi} = 6.5$. \textbf{(e)} Measured signal-to noise ratio $C_{spin}/\delta C_{\pi}$ (magenta dots) as a function of the measurement time $t_m$, and fit with the function $A \sqrt{t_m}$ (solid line). Data taken at $B_0=421.042$ mT and $\theta=-0.024^{\circ}$.}
\label{fig4}
\end{figure}

To demonstrate the stability and reproducibility of the peaks, we perform a two-dimensional magnetic field scan by recording a background-corrected average number of counts (see Methods), named $\langle \tilde{C} \rangle$ hereafter, as a function of $B_0$ and $\theta$ (see Fig.~\ref{fig3}). Eight different spin peaks are resolved, and their spectrum is readily followed in magnetic field. It appears that each ion has its own gyromagnetic tensor $\mathbf{\gamma}$, close to $\mathbf{\gamma}_0$ but with different values for the principal axes and also a symmetry axis that can slightly deviate from the $c$-axis, vividly illustrating the concept of inhomogeneous broadening. The lines are so narrow that each ion $\mathbf{\gamma}$ could, in principle, be determined to better than $10^{-6}$ accuracy (using a suitably calibrated magnetic field). Because the deviation $\delta \mathbf{\gamma}$ of the gyromagnetic tensor from the ensemble-averaged $\mathbf{\gamma}_0$ is due to the local electrostatic and strain environment, its accurate measurement can also be turned into a sensitive way to probe it (as done with NV centers in diamond in particular~\cite{broadway_microscopic_2019}). Note that our measurements also call for a better modeling of the response of rare-earth ion spins to applied electric or strain fields. 

We now select one of the peaks ($s0$) and bring further proof of its single-spin nature. We first measure the excess counts $\langle \tilde{C} \rangle$ as a function of the microwave pulse duration, and observe sinusoidal oscillations with a frequency that depends linearly on the pulse amplitude (see Fig.~\ref{fig4}a and b), as expected for the Rabi oscillation of a single spin. Superposed on these oscillations is a gradual increase in counts, which we attribute to heating of the bath of defects that are responsible for the resonator internal losses upon microwave excitation, as observed in~\cite{albertinale_detecting_2021} (see Methods). We then use the SMPD to measure the photon statistics of the fluorescence signal and reveal its single emitter nature. For this task, we acquire a large number of fluorescence traces following a $\pi$ pulse, label them by index $j$, and compute the dark-count-corrected intensity-intensity correlation function $g^{(2)}(k)$ between traces whose index differs by $k$ (see Methods). For $N$ emitters, $g^{(2)}(0)$ should be equal to $(N-1)/N$; in particular, $g^{(2)}(0)$ should be equal to $0$ for a single-emitter since it can emit only one photon in each sequence. We measure $g^{(2)}(0) = 0.23 \pm 0.06$, and $g^{(2)}(k) = 1 \pm 0.04$ for $k\neq 0$ (see Fig.~\ref{fig4}c), thus showing clear anti-bunching in each sequence, whereas the emission from different sequences is uncorrelated. The non-zero value of $g^{(2)}(0)$ may be due to heating; in any case, the fact that its value is well below $0.5$ further suggests that the spectral peak under study is a single microwave photon emitter, in the form of an individual $\mathrm{Er}^{3+}$ electron-spin. 

We use the same dataset to quantify the single-spin SNR for a certain measurement time $t_m$. For that, we sum the counts obtained over sequences played during a $t_m$ time window, integrated over the first $2$\,ms following the excitation pulse, yielding the number of counts $C$. Figure ~\ref{fig4}d shows the count probability histogram $p(C)$ for $t_m=1$\,s, with and without $\pi$ pulses applied. Both histograms are well reproduced by a Poissonian distribution (see Fig.~\ref{fig4}d). The single-spin SNR defined as $C_{spin}/\delta C_\pi$ has a value of $1.91$. Here, $C_{spin}$ is the difference between the mean number of counts and $\delta C_\pi$ the half-width of the distribution with $\pi$ pulse applied. A comparison with the expected SNR requires measuring the overall efficiency $\eta$, which we find to be equal to $\eta = 0.12 \pm 0.01$ by integrating the fluorescence signal after the $\pi$ pulse with subtracted background. This value of $\eta$ results from the SMPD finite efficiency, resonator internal losses, and microwave losses in-between the spin resonator and the SMPD (see Methods). We deduce an optimal theoretical SNR of $\sim 2.5$ (see Methods), close to our measured value. We also verify that the SNR scales as the square root of the measurement time $t_m$ up to at least 1 minute (see Fig.~\ref{fig4}e), indicative of good measurement stability.

\begin{figure}
\centerline{\includegraphics[width=0.5\textwidth]{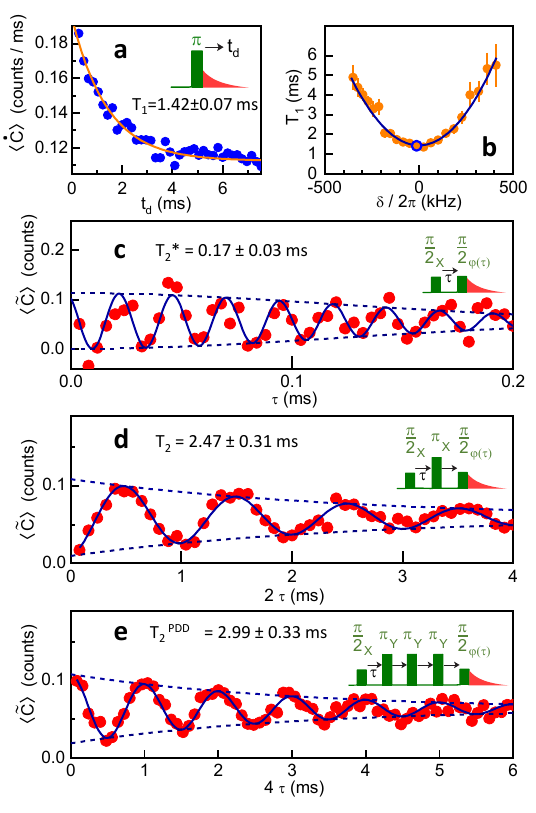}}
\caption{
\textbf{Coherence time measurements of spin $s6$.}
\textbf{(a)} Energy relaxation: measured average count rate $\langle \Dot{C}\rangle$ (blue dots) as a function of delay $t_d$ after a resonant $\pi$ excitation pulse. Exponential fit (solid orange line) yields the energy relaxation time $T_1$. \textbf{(b)} Purcell effect: measured $T_1$ as a function of spin-resonator frequency detuning $\delta$ (orange dots). A fit to $\Gamma_R^{-1}(\delta)$ (solid black line) yields the spin-resonator coupling constant $g_0/2\pi=3.54\pm0.15$ kHz. \textbf{(c)} Ramsey sequence (see inset): measured excess counts $\langle \widetilde{C}\rangle$ versus delay time $\tau$ between two resonant $\pi/2$ pulses with relative phase $\varphi(\tau) = 2 \pi \Delta \tau$ and $\Delta = 0.025$\,MHz (dots). The corresponding fit (solid line) by a sine function with a Gaussian-decaying envelope (dash lines) yields a coherence time $T_2^* = 0.17 \pm 0.03$\,ms. \textbf{(d)} Hahn-echo sequence (see inset): measured excess counts $\langle \widetilde{C}\rangle$ versus delay $\tau$ between subsequent pulses with a linearly increasing phase $\varphi(\tau) = 2 \pi \Delta \tau$ with $\Delta = 0.001$\,MHz on the last pulse (red dots). The corresponding fit and its envelope (solid and dash lines) yield a coherence time $T_2 = 2.47\pm 0.31$\,ms. \textbf{(e)} Periodic Dynamical Decoupling sequence (see inset): measured excess counts $\langle \widetilde{C}\rangle$ versus inter-pulse delay time $\tau$ (red dots). A linearly increasing phase $\varphi(\tau) = 2 \pi \Delta \tau$ with $\Delta = 0.001$\,MHz is imparted on the last pulse. Corresponding fit and its envelope (solid and dash lines) are shown, yielding the coherence time $T_2^{PDD} = 2.99\pm 0.03$\,ms. Data taken at $B_0 = 422.085$ mT and $\theta=-0.003^{\circ}$.}
\label{fig5}
\end{figure}

The ability to address individual spins with microwaves opens the way to using them as spin qubits for quantum computing, and it is thus interesting to characterize their coherence properties. The longitudinal relaxation time $T_1$ is obtained simply from the fluorescence curve decay; we select a spin ($s6$) with $T_1 = 1.46 \pm 0.05$\,ms (see Fig.~\ref{fig5}a) at resonance ($\delta = 0$). We then measure the free-induction-decay time using a Ramsey sequence $\pi/2_X - \tau - \pi/2_\Phi$, with the relative inter-pulse phase $\Phi = 2 \pi \Delta \tau$, where $\Delta = 0.025$\,MHz. The excess count $\langle \tilde{C} \rangle$ shows oscillations at frequency $\Delta + \delta$, damped with an approximately Gaussian shape and a characteristic relaxation time $T_2^* = 170 \pm 33\,\mu \mathrm{s}$, corresponding to a $\sim 2$\,kHz single-spin linewidth (see Fig.~\ref{fig5}c). We use the Ramsey sequence to accurately measure $\delta$, making it possible to determine the dependence of the spin longitudinal relaxation time $T_1$ on $\delta$ (Fig.~\ref{fig5}b). It is seen to increase with $\delta$ quadratically, in agreement with the expected $\Gamma_R^{-1}$ dependence~\cite{bienfait_controlling_2016}; a fit yields a coupling constant $g_0/2\pi = 3.56$\,kHz (see Fig.~\ref{fig5}b). This confirms that non-radiative relaxation is negligible in our measurements (see Methods), and that $T_1 \simeq \Gamma_R^{-1}$ for the most strongly coupled spins. 

The Hahn echo coherence time is measured by applying the sequence $\pi/2_X - \tau - \pi_Y - \tau - \pi/2_\Phi$~\cite{billaud_microwave_2022}, with $\Phi = 2 \pi \Delta \tau$. An oscillation at frequency $\Delta$ is observed in $\langle \tilde{C} \rangle$, exponentially relaxing with a characteristic time $T_2 = 2.47 \pm 0.31$\,ms. This is close to the radiative decay limit $2T_1$, indicating that the pure dephasing echo contribution is $\sim 16 \pm 5$\,ms, in line with measurements on ensembles of $\mathrm{Er}^{3+}:\mathrm{CaWO}_4$ electron spins~\cite{le_dantec_twenty-three-millisecond_nodate}. This dephasing can be suppressed further by a 3-$\pi$-pulse Periodic Dynamical Decoupling sequence, yielding a transverse relaxation time $T_{2}^{PDD} = 2.99 \pm 0.33$\,ms, which is equal to $2T_1$ to the accuracy of the measurement. These coherence times were also measured on a set of five $\mathrm{Er}^{3+}$ electron spins; $T_2^*$ varies strongly among these ions (between $5 \mu \mathrm{s}$ and $300 \mu \mathrm{s}$), whereas $T_2$ and $T_{2}^{PDD}$ are consistently close to $2T_1$ (see Methods). The variation of coherence time among different spins can be explained by the varying nuclear spin or paramagnetic environment of each ion, and also possibly their degree of exposure to surface magnetic noise given their approximate depth of $\sim  100-150$\,nm according to Fig.~\ref{fig2}~\cite{myers_probing_2014,ranjan_spatially-resolved_2021}. It is also noteworthy that the coherence times measured here are on par with the longest reported for individual electron spins in solid-state~\cite{muhonen_storing_2014}, in a platform which gives access to several tens of these spin qubits by simply tuning the magnetic field.

We now discuss the significance of our results for practical single electron spin resonance spectroscopy. One particularly interesting aspect of our method is its applicability to a broad range of paramagnetic species, provided their radiative relaxation rate $\Gamma_R$ can be enhanced up to $\sim 10^3 \mathrm{s}^{-1}$ or higher by the Purcell effect, and their non-radiative relaxation rate is smaller than $\Gamma_R$. Note that no requirement on the coherence time applies, as the fluorescence signal is entirely incoherent. Indeed, many paramagnetic impurities have non-radiative relaxation rates in the range of $10^{-3} - 10^3\,\mathrm{s}^{-1}$ at $\sim 1-4$\,K~\cite{castle_resonance_1965,gayda_temperature_1979,zhou_electron_1999}, and thus also likely at millikelvin temperatures. Although reaching the desired radiative relaxation rate of $\Gamma_R > 10^3 \mathrm{s}^{-1}$ was made easier in this work by the large transverse $g$-factor of $8.3$ in $\mathrm{Er}^{3+}:\mathrm{CaWO}_4$, this large relaxation rate was also demonstrated for donor spins in silicon with $g$-factors of only $2$, using a similar resonator geometry but with a narrower and shorter wire~\cite{ranjan_electron_2020}. Whereas in our experiment the spins are located in the sample supporting the resonator, it is also possible to deposit a small volume of a spin-containing insulating material, such as a powder or micro-crystal, onto a pre-fabricated resonator device. Such an approach could be suitable for measuring individual rare-earth-ion-containing molecules~\cite{vincent_electronic_2012}, nanocrystals~\cite{casabone_dynamic_2021}, or proteins whose active center contains a transition-metal-ion~\cite{coremans_w-band_1994,doorslaer_strength_2007}. Based on the $~10\,\mu \mathrm{m}^3$ detection volume demonstrated here using $\mathrm{Er}^{3+}:\mathrm{CaWO}_4$, we extrapolate that a $~0.5\,\mu \mathrm{m}^3$ detection volume would be achievable for an electron-spin with a g-factor of two, under the same experimental conditions. All these metrics could be improved with better SMPD performance, in particular reduced dark count rates, highlighting a strong motivation for the continued development of SMPD devices. 

In conclusion, we report spectroscopic measurements of single rare-earth-ion electron spins by detecting their microwave fluorescence, gaining four orders of magnitude in spectral resolution by resolving the ensemble inhomogeneous linewidth. In our experiment, tens of individual spins with coherence times in excess of 1 millisecond are interfaced with the same microwave resonator, which opens new perspectives for hybrid quantum computing. Because of its broad applicability, large detection volume, and spectroscopic capability, our detection method comes close to practical single electron spin resonance at cryogenic temperatures, and may thus open new applications in ESR spectroscopy.

\subsection*{Acknowledgements}
{We acknowledge technical support from P.~S\'enat, D. Duet, P.-F.~Orfila and S.~Delprat, and are grateful for fruitful discussions within the Quantronics group. We acknowledge support from the Agence Nationale de la Recherche (ANR) through the Chaire Industrielle NASNIQ under contract ANR-17-CHIN-0001 cofunded by Atos, and through the MIRESPIN (ANR-19-CE47-0011) and DARKWADOR (ANR-19-CE47-0004) projects. We acknowledge support of the R\'egion Ile-de-France through the DIM SIRTEQ (REIMIC project), from CEA through the DRF-Impulsion porgram (grant RPENANO), from the AIDAS virtual joint laboratory, and from the France 2030 plan under the ANR-22-PETQ-0003 grant. This project has received funding from the European Union Horizon 2020 research and innovation program under ERC-2021-STG grant agreement no. 101042315 (INGENIOUS) and Marie Sklodowska-Curie grant agreement no. 792727 (SMERC). Z.W. acknowledges financial support from the Sherbrooke Quantum Institute, from the International Doctoral Action of Paris-Saclay IDEX, and from the IRL-Quantum Frontiers Lab. S.B. thanks the support of the CNRS research infrastructure INFRANALYTICS (FR 2054). We acknowledge IARPA and Lincoln Labs for providing the Josephson Traveling-Wave Parametric Amplifier.}

\subsection*{Author contributions}
{A.F. and P.G. grew the crystal, which M.L.D., Z.W. and S.B. characterized through CW and pulse EPR measurements. Z.W., D.V., P.B., E.F. designed the spin resonator. Z.W. fabricated the spin resonator. L.B., E.F. designed the SMPD. L.B. fabricated the SMPD. M.R. designed and installed the magnetic field stabilization. Z.W., L.B., E.F. took the measurements. Z.W., P.B., E.F. analyzed the data. Z.W., P.B., D.V., E.F. wrote the article, with contributions from all the authors. P.B. and E.F. supervised the project.}

\bibliographystyle{naturemag}
\bibliography{main}

\clearpage

\end{document}



\title{Method}

\date{\today}

\maketitle
\subsection{Sample}

\begin{figure}[!tbh]
\includegraphics[width=0.40\textwidth]{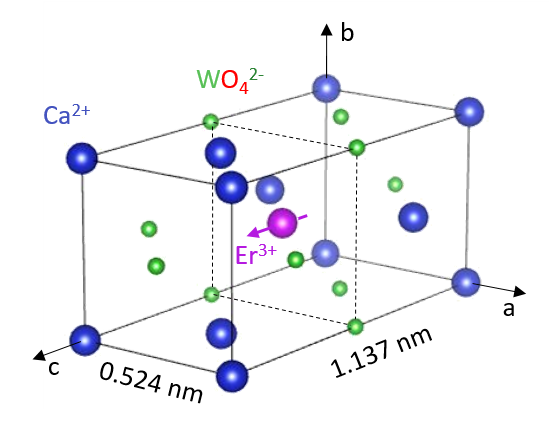}
\caption{
\textbf{Crystal structure of $\mathrm{Er^{3+}:CaWO_4}$} (oxygen is not shown for clarity).
}
\label{figS0}
\end{figure}

The $\mathrm{CaWO_4}$ crystal used in this experiment originates from a boule grown by the Czochralski method from $\mathrm{CaCO_3}$ (99.95\% purity) and $\mathrm{WO_3}$ (99.9 \% purity). A sample was cut in a rectangular slab shape ($7~\mathrm{mm}\times4~\mathrm{mm}\times0.5~\mathrm{mm}$), with the surface approximately in the $(ac)$ crystallographic plane, and the $c$-axis parallel to its short edge. The crystal structure is tetragonal with unit cell constants $a=b=0.524\,nm$ and $c=1.137\,nm$, as shown in Extended Data Fig.\,\ref{figS0}. The erbium ions $\mathrm{Er}^{3+}$ substitute to the calcium ions $\mathrm{Ca}^{2+}$ (with long-range charge compensation in the crystal). These sites have a $S_4$ symmetry, leading to a gyromagnetic tensor with only diagonal elements in the $(a,b,c)$ plane $\gamma_a = \gamma_b \equiv \gamma_\perp = 2\pi \times 117.3$\,GHz/T, and $\gamma_{c} \equiv \gamma_{||} = 2\pi \times 17.45$\,GHz/T \cite{antipin_a._paramagnetic_1968}. The residual doping concentration of erbium is 3.1 $\pm$ 0.2 ppb, measured from continuous-wave EPR spectroscopy \cite{le_dantec_electron_2022}. 


\begin{figure}[!tbh]
\includegraphics[width=0.45\textwidth]{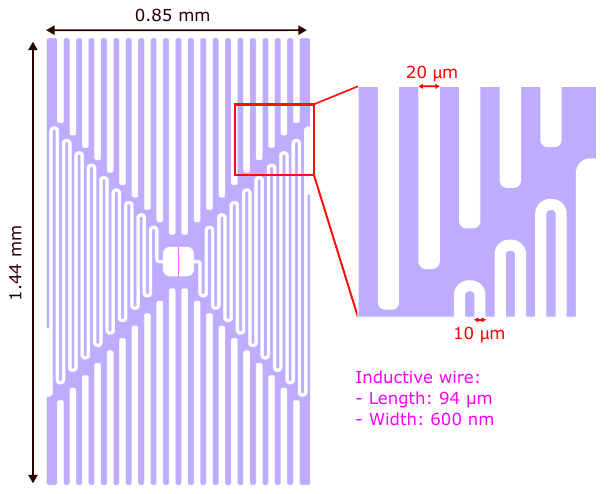}
\caption{
\textbf{Resonator design.} 
}
\label{figS1}
\end{figure}

On top of this sample, a lumped-element LC resonator was fabricated by sputtering 50nm of niobium and patterning the thin film by electron-beam lithography and reactive ion etching. The sample is placed in a 3D copper cavity with a single microwave antena and SMA port used both for the excitation and the readout in reflection. As shown in Extended Data Fig.\ref{figS1}, the "bow-tie" shaped resonator consists of an interdigitated capacitor shunted by a $94~\mathrm{\mu m} \times 600~\mathrm{nm}$ inductive wire in the middle. From finite-element microwave simulations, we deduce an impedance of 17.5\,$\Omega$. The geometric inductance of the inductance wire contributes 33\% of the total resonator inductance. The top and bottom capacitor pads are shaped as parallel fingers in an effort to improve the resonator resilience to an applied residual magnetic field perpendicular to the metallic film (along x).

\subsection{Experimental setup}
The complete setup schematic is shown in Extended Data Fig. \ref{sub_fig1}
\subsubsection{Room-temperature setup}
Its room-temperature part uses five microwave sources and one FPGA-based instrument (OPX platform from Quantum Machine) for arbitrary waveform generation, digitization, and real time feedback. The OPX instrument contains 10 channels of analog outputs (AO), 10 digital outputs (DO) and 2 analog inputs (AI).

The pulses used to drive the spins are generated by I/Q mixing a pair of in-phase (I) and quadrature (Q) signals from the OPX with a local oscillator (LO - orange) at the spin resonator frequency $\omega_0$. The upconverted microwave signal is then split over 2 branches, one of them including an about 40 dB amplifier, which are then recombined. Only one of the branch is chosen to propagate the signal, with two fast switches controlled by digital lines from the OPX. The spin excitation pulses enters the dilution refrigerator through line 2.

The SMPD operation (see \cite{albertinale_detecting_2021} for details) involves one dc-current and three microwave sources, the role of which are as follows: (1) A Yokogawa current source (red) provides the necessary flux bias to bring the SQUID-tunable buffer resonator of the SMPD at $\omega_b$ in resonance with the spin resonator at $\omega_0$, so that the fluorescence photons emitted by the spins are at the center of the SMPD detection bandwidth. (2) A pump drive (purple) at frequency $\omega_p$ enables a four-wave mixing process converting an incoming photon in the buffer into an excitation of a superconducting transmon qubit at $\omega_q$ and a photon in a readout (waste) resonator at $\omega_w$, according to $\omega_p+\omega_b=\omega_q+\omega_w$. (3) The readout of the qubit is performed by probing by homodyne detection (green) the qubit-state dependent dispersive shift of the readout resonator. (4) Control pulses of the qubit are generated with a sideband mixer from one OPX IF output and the blue LO source. They are combined with the pump pulse and are sent to line 7.

A Rohde \& Schwarz microwave source (yellow) at the input of line 5 provides the pump power for a traveling wave parametric amplifier (TWPA) placed at 10 mK.

\subsubsection{Low-temperature setup}

The spin excitation pulses (line 2) are heavily attenuated ($\sim 110\, \mathrm{dB}$) to minimize the thermal excitation of the qubit and dark counts. They are directed,  through a double- and a single-junction circulator, to the antenna of the cavity containing the spin resonator. The reflected and output signals on this antenna are routed to the input of the SMPD through a single-junction circulator. To pre-characterize the spin resonator as well as the SMPD, the signal reflected on the SMPD input is routed to room-temperature via the same single- and double-junction circulators and output line 1 with a first HEMT amplifier; isolation of the experiment from this HEMT is provided by a double circulator and an extra 10dB attenuation. Note that during all measurements reported in the main text, this line 1 was left open and its HEMT switched off.

SMPD qubit readout pulses are sent via the attenuated line 4 and a double circulator. The reflected signal is routed to a Josephson Traveling Wave Parametric Amplifier (TWPA) pumped from line 5 via a directional coupler and to a second HEMT at the 4 K stage. A double-circulator isolates the TWPA from this HEMT.

The SMPD pump tone and qubit reset pulses are applied via line 7 and its 20dB directional coupler. The other 2 ports of the coupler are connected to a 50$\Omega$ load at 800 mK and a 50$\Omega$-loaded circulator at 10mK, in order to minimize the noise induced by the dissipation of these signals. 

\subsection{Single microwave photon detector}

The SMPD is operated in cycles of $12.8\ \mathrm{\mu s}$ on average. Each cycle is composed of three steps: (i) the pumped conversion of an incoming photon into a qubit excitation during $10\ \mathrm{\mu s}$, (ii) the qubit dispersive readout lasting $2\ \mathrm{\mu s}$, and (iii) the conditional reset of the qubit to its ground state if it was detected excited. This reset consists of one or several $\pi$-pulse(s) applied to the qubit until it is measured in its ground state. The conditional reset time is thus non-deterministic, and lasts from $0.7\ \mathrm{\mu s}$ (feedback time with the OPX) to $0.7 + (2+0.7)k \ \mathrm{\mu s}$, with $k$ the number of $\pi$ pulses applied.

At each cycle, a count $C=1$ is detected if the qubit is found in its excited state (before the reset), and the count time is recorded with sub-microsecond accuracy.

The SMPD is characterized independently, in absence of spin signal, by measuring its key figures of merit in terms of dark count rate, efficiency, and bandwidth.

\textit{Dark count rate.} - We define dark counts as the counts that are not due to the spins, that is those originating from spurious excitation of the transmon qubit in absence of incoming photons, and those due to unwanted photons present at the SMPD input \cite{albertinale_detecting_2021}. For this detector, a dark count rate of  $106 \pm 3\ s^{-1}$ has been measured over 24 hours  (data from Extended Data Fig. \ref{fig_g2}). We observed slow darkcount rate fluctuations over week time scales ranging typically from $130\,s^{-1}$ to $90\,s^{-1}$ mainly due to variation in qubit $T_1$ and a slow cooling down of the line and of the qubit. We can discriminate dark count contributions from the microwave line thermal occupancy, from the pump heating and from the qubit thermal occupancy by switching off and detuning the pump tone from the four wave mixing frequency. By detuning the pump by $10\ \mathrm{MHz}$, the detection efficiency is set close to zero but the pump heating load persists, we measure count rates of 11 $s^{-1}$. When the pump is turned off, the dark count rate is 9 $s^{-1}$, indicating that the pump heating is negligible. From these measurement, we conclude that 92\% of the dark counts come from thermal microwave photons reaching the SMPD via its input line. This corresponds to a thermal population of $\sim 2.4 \times 10^{-4}$ photons in the line, and to an effective temperature of $\sim 42$\,mK, to be compared to the measured $10$\,mK base temperature of the refrigerator.

\begin{figure}[!tbh]
\includegraphics[width=0.45\textwidth]{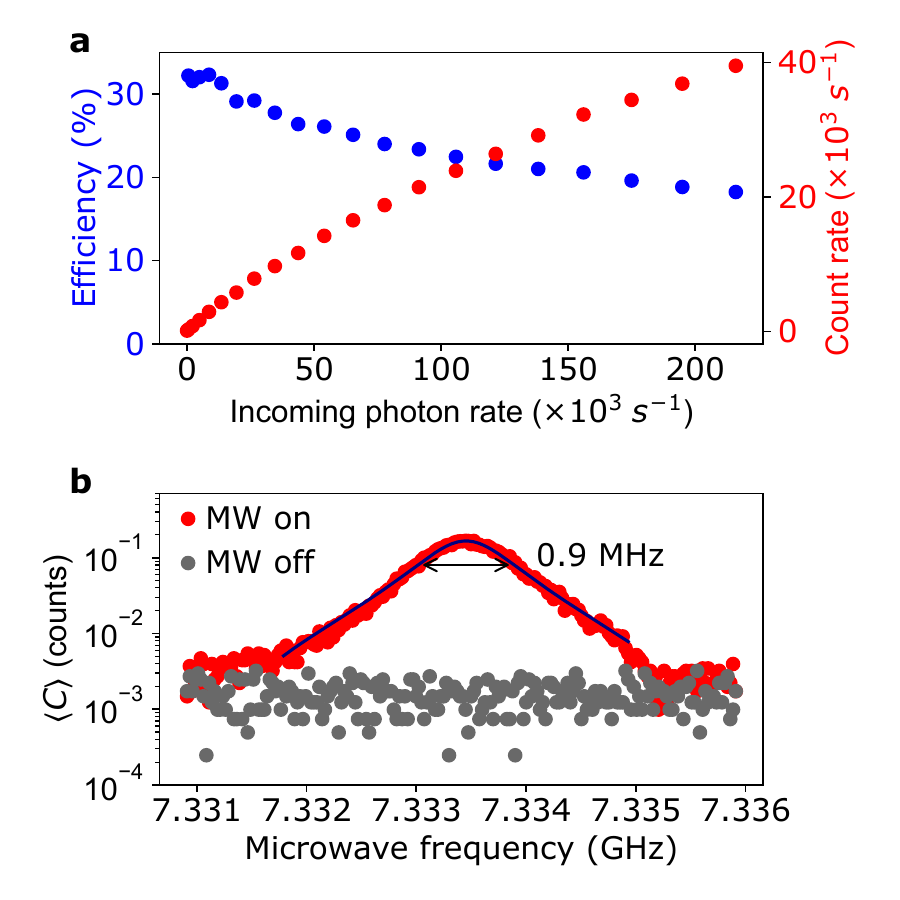}
\caption{
\textbf{SMPD characteristics.} 
\textbf{(a)}  SMPD efficiency. Detected click rate (red) and efficiency (blue) as a function of input photon flux. Below $10^4$ $s^{-1}$ (linear regime), an efficiency of 32\% is obtained. \textbf{(b)} SMPD bandwidth. Average number of detected counts as a function of photon frequency when the input microwave tone is switched on (red dots) or off (gray dots). The solid line is a Lorentzian fit to the data yielding a FWHM bandwidth of 0.9 MHz.}
\label{figSMPD}
\end{figure}

\textit{Efficiency.} - The detector efficiency is measured by shining a microwave tone of known power at the detector input. The average input photon flux for a given applied power is calibrated in-situ by measuring the transmon qubit a.c. Stark shift and dephasing \cite{gambetta_qubit-photon_2006}. The SMPD efficiency is then simply taken as the ratio between the counts detected over 1s and the photon flux (in photon/s). It was measured for different input powers, as shown in Extended Data Fig.\ref{figSMPD}a:  At hundreds of input photons per second, a value close to the fluorescence signal obtained at high excitation power, the efficiency is $\eta_{SMPD} = 0.32$. The detector saturates and the efficiency drops at input fluxes above $10^4$ photons/s. Further optimizing the lifetime of the transmon qubit as well as the readout and pump power for four-wave mixing, would probably yield a better efficiency.

\textit{Bandwidth.} - The detector bandwidth is extracted by measuring the average detected counts $\langle C \rangle$ as a function of the microwave frequency. Each $\sim 10 \mu s$-long pulse contains 0.5 photon on average to mimic single spin detection. The full width at half maximum (FWHM) of a Lorentzian fit gives a bandwidth of 0.9 MHz for the detector, as shown in Extended Data Fig. \ref{figSMPD}b.

\subsection{Average number of counts}
In the case of Fig 2, We calculate the average number of counts
\begin{equation}
    \langle C \rangle = \frac{1}{N} \sum_{n=1}^{N}\sum_{0}^{T} c_{n}(t_d)
\end{equation}
by summing the counts from $t_d=0$ to $T$, with $N$ the number of repetitions of the experiment and $c_n(t_d)$ the 0 or 1 SMPD outcome at time $t_d$.

For the other figures (Figs.\,3, 4 and 5) involving single spins, the fluorescence signal is measured as a function of time up to $\sim 5T_1$ after the excitation pulse, and its second half (close to the background level) is subtracted from the first part, leading to a background-corrected average number of counts  
\begin{equation}
    \langle \tilde C \rangle = \frac{1}{N} \sum_{n=1}^{N}\left[\sum_0^{T/2} c_{n}(t_d)-\sum_{T/2}^{T} c_{n}(t_d)\right].
\end{equation}

\subsection{Magnetic field alignment and stabilization}

\begin{figure}[!tbh]
\includegraphics[width=0.45\textwidth]{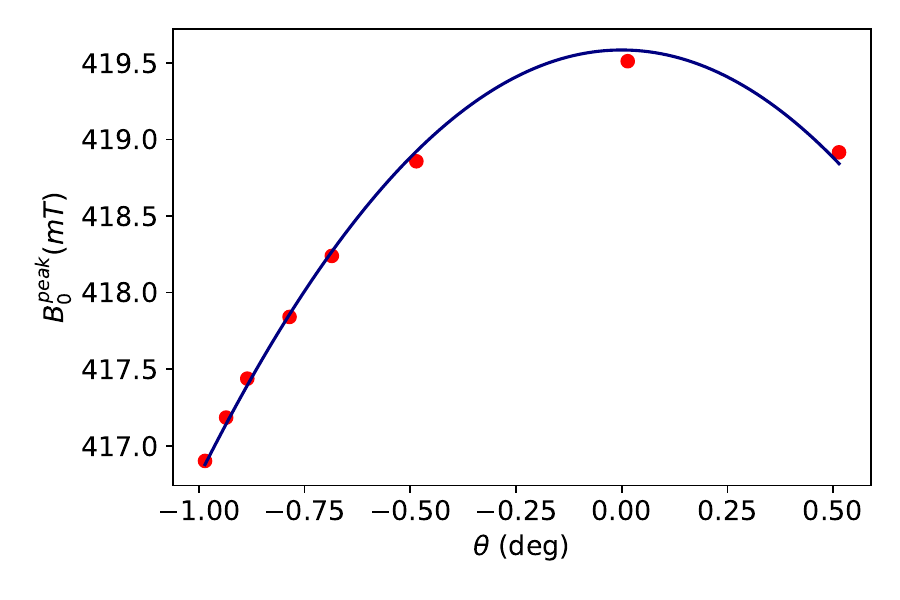}
\caption{
\textbf{Magnetic field alignement.}
Measured (dots) magnetic field  $B_0^{peak}$ at which the center of the spin ensemble line is found, as a function of the angle $\theta$ that the field makes with the $c$ axis projection. The fit with eq. \ref{eq:field_alignment} (line) to the data yields the $\theta=0$ origin (see text), as well as the angle between the $c$ axis and the sample plane, $\beta = 0.5^\circ$.
}
\label{fig_field_alignment}
\end{figure}

The magnetic field $B_0$ is generated by a 1T/1T/1T $3$-axis superconducting vector magnet. Magnetic field alignment proceeds in two steps. We first align the field in the sample plane ($y-z$) by applying a small field of $50$\,mT, and minimizing the resonator losses and frequency shift with respect to the zero-field values~\cite{le_dantec_electron_2022}. We then determine the direction of the projection of the crystallographic $c$-axis on the sample plane, defined as $\theta = 0^\circ$, by measuring the erbium ensemble line (as shown in Fig.2 of the main text) for various angles in the ($y-z$) plane. Due to the anisotropy of the gyromagnetic tensor $\gamma_0$, $B_0$ can be expressed with angle $\theta$ and $\beta$ as
\begin{equation}
    B_0^{peak}=\hbar\omega_0/\sqrt{\gamma_{\parallel}^2cos^2\theta cos^2\beta+\gamma_{\perp}^2(1-cos^2\theta cos^2\beta)},
\label{eq:field_alignment}
\end{equation}
where $\theta= 0^\circ$ corresponds to the maximum of $B_0$ and $\beta$ is the angle between the c-axis and sample plane. As shown in Extended Data Fig. \ref{fig_field_alignment}, the fitting (solid line) with eq. \ref{eq:field_alignment} to the data (dots) yields $\beta=0.5^{\circ}$.

Note that this procedure does not guarantee that the resonator nanowire exactly coincides with the $\theta =  0^\circ$ direction determined by our alignment procedure. In fact, a small residual angle (possibly of order $1^\circ$) likely exists between the two directions. Since we have no way to determine this angle, and since its non-zero value has negligible impact on any of the results found in the article, we used a zero value by simplicity for plotting Fig. 2a.


The stability of the magnetic field is determined by the mode of operation (current supplied mode or persistent mode) of the three superconducting coils of the vector magnet and by their current sources. For the spectroscopy data of the main text (Figs 2 and 3), the current supplied mode is used with a commercial current source (Four-Quadrant Power Supply Model 4Q06125PS from AMI). On the contrary, the data of Figs. 4 and 5 require to tune the spin-resonator frequency difference $\delta$ and to keep it stable (less than $10 \mathrm{kHz}$ variation) over long periods of time. To achieve this goal, we use the fact that one of our coils is nearly aligned with the $z$-axis and thus provides the largest component of the $B_0$ field; we thus minimize the noise by placing it in persistent mode. Then, the coil closest to the $y$ axis is used to fine-tune $\delta$. The (much smaller) current through that coil is moreover further stabilized using a custom-made feedback loop based on a current meter (Keithley $2700$ model).

\subsection{Microwave induced heating and corresponding spurious signal}
We now discuss heating effects observed (as in~\cite{albertinale_detecting_2021}) after a microwave pulse is applied to the spin resonator. To evidence and clarify this point, we first measure in three different cases the transcient signal recorded after an excitation pulse resonant with the SMPD buffer resonator:


\begin{enumerate}
\item Normal operation on a single spin: single spin, spin resonator and SMPD buffer are all in resonance.
\item Normal operation in absence of spins: All spins are far off-resonance, and spin resonator and SMPD buffer are on resonance.
\item Complementary diagnosis: All spins, spin resonator and SMPD buffer are detuned from one another.
\end{enumerate}

A $6\mu \mathrm{s}$-long excitation pulse is applied at time $t=0$; the photon counting sequence starts 1 ms before the pulse, is interrupted (SMPD switched off) during the pulse duration, and is restarded during several ms. 

\begin{figure}[!tbh]
\includegraphics[width=0.45\textwidth]{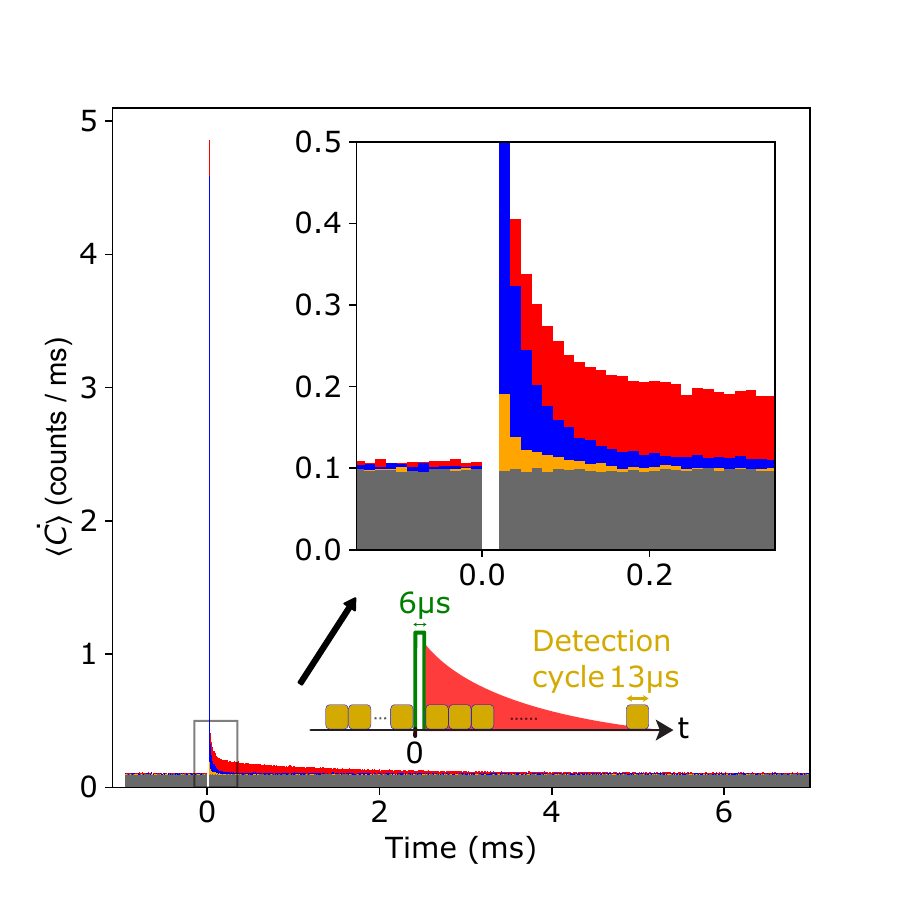}
\caption{
\textbf{Transient response of the system after microwave excitation.}
Measured average click rate versus time before and after a 6 $\mu s$-long microwave excitation pulse is applied at time 0, for cases 1 (red), 2 (blue) and 3 (orange) - see text. SMPD is switched off during excitation. Dark count background is indicated in grey. Inset is a zoomed-in view around time 0. 
}
\label{figHeating_pulse}
\end{figure}

In normal operation (case 1 and 2 - red and blue in Fig. \ref{figHeating_pulse}), a count rate spike is observed in the bin immediately following the excitation pulse; it corresponds to the decay (at rate $\kappa^{-1}$) of the microwave energy stored by the pulse in the spin resonator. This spike disappears when the spin resonator is detuned from the signal (case 3), as expected. After the spike, even when no spin signal is present (case 2 - blue), extra counts above the background (grey) are however observed over a time window of about $0.3$\,ms after the pulse, with a decay time of $\sim 100 \mu \mathrm{s}$. This extra signal is reminiscent of the one observed over $\sim 10$\,ms in~\cite{albertinale_detecting_2021}, possibly shorter in the present work due to lower excitation pulse powers. For comparison, when detuning the spin resonator from the excitation (case 3 - orange in Fig. \ref{figHeating_pulse}), the extra count rate is lower and reaches the background steady-state much faster. 
All these measurements with no spins indicate that the spurious extra counts decaying over $\sim 100 \mu \mathrm{s}$ in normal operation originate from the excitation and subsequent radiative decay of systems that are resonantly coupled to the spin resonator. It is tempting to identify them with the two-level-system bath that causes field decay and phase noise in superconducting circuits. In normal operation with a spin (case1 - red in Fig. \ref{figHeating_pulse}), these spurious extra counts of course adds to the relevant signal coming from the spin. To lower the impact of this transient heating effect, the results of Fig. 4c (resp. Figs. 5a and b) were obtained  by discarding the counts detected in the first $100 \mu \mathrm{s}$ (resp. $50 \mu \mathrm{s}$) time window following the excitation.

We finally study the dependence of this heating effect in absence of resonant spins on the excitation pulse duration and amplitude. For that we integrate over $8$\,ms the number of counts $\langle C \rangle$ after an excitation pulse, and repeat the sequence every $8$\,ms. The results in Extended Fig.4 show an increase of $\langle C \rangle$ as a function of pulse duration and amplitude, as well as a characteristic time for this increase with pulse duration that decreases as excitation amplitude increases. We also verified that this increase of $\langle C \rangle$ is not due to microwave heating of the line attenuators, by repeating the same measurements with the  spin resonator detuned from the SMPD buffer: a much smaller effect is observed, indicating that the excess counts do come from the spin resonator.

\begin{figure}[!tbh]
\includegraphics[width=0.45\textwidth]{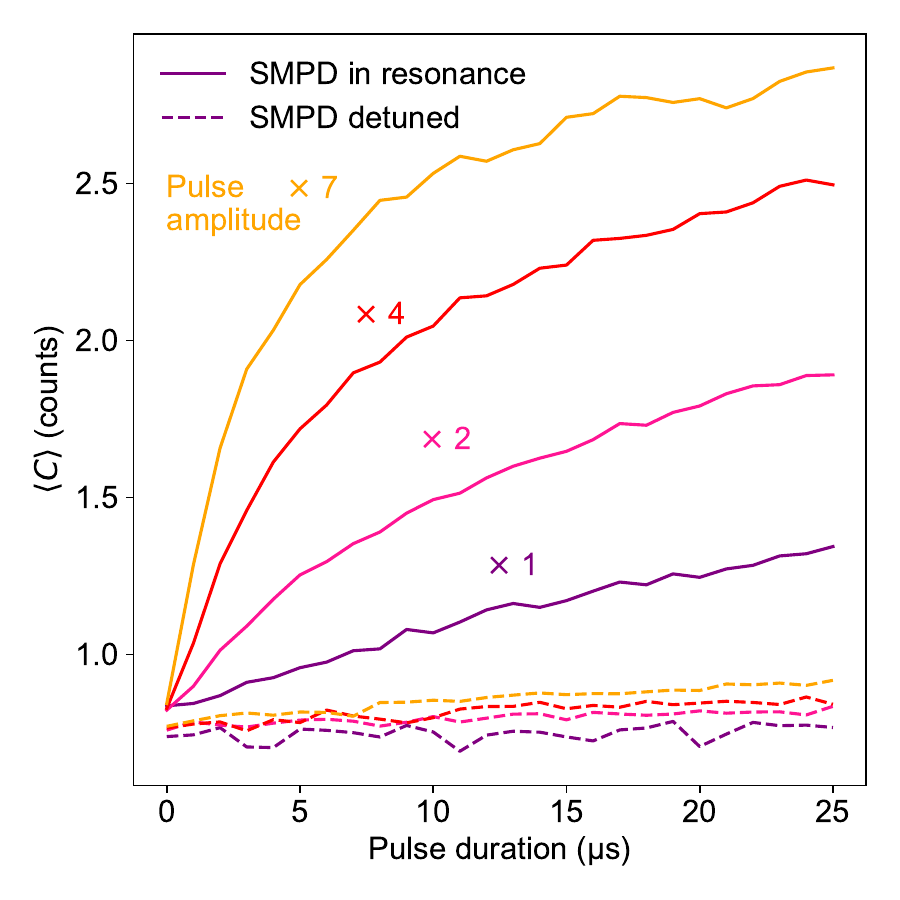}
\caption{
\textbf{Heating versus spin excitation duration and amplitude.}
Measured average counts integrated over a 8ms-long window after an excitation pulse as a function of pulse duration and amplitude A, obtained when the spin resonator is detuned from (dash line) or in resonance with (solid line) the SMPD buffer. The excitation pulse frequency is always tuned to the SMPD buffer one.
}
\label{figHeating_rabi}
\end{figure}

\subsection{Intensity-intensity correlation measurements}

We now provide more details on the intensity-intensity correlation measurements used to prove the single spin character of our experiment. 

The dataset to be analyzed corresponds to two series of 4363635 sequences labelled from i=0 to i=4363634 repeated every $t_r = 7.5$\,ms, where one serie includes a $\pi$ pulse at time $t=0$ and the other has no excitation pulse. Time $t=0$ is followed by $600$ SMPD cycles. As explained in the Heating section, the first $100 \mu \mathrm{s}$ window after the excitation pulse is excluded from the analysis in order to minimize the impact of the heating effect. The count data in the rest of the sequence are then grouped in subsequent $350 \mu \mathrm{s}$-long timebins indexed by j (with j running from 0 to 20), and centered at time $\tau_j = 100 + (2j+1) \times 350/2~\mu \mathrm{s}$. The corresponding number of counts in the bin $j$ of sequence $i$ is denoted as $n^{(i)}_j$.

\begin{figure}[!tbh]
\includegraphics[width=0.45\textwidth]{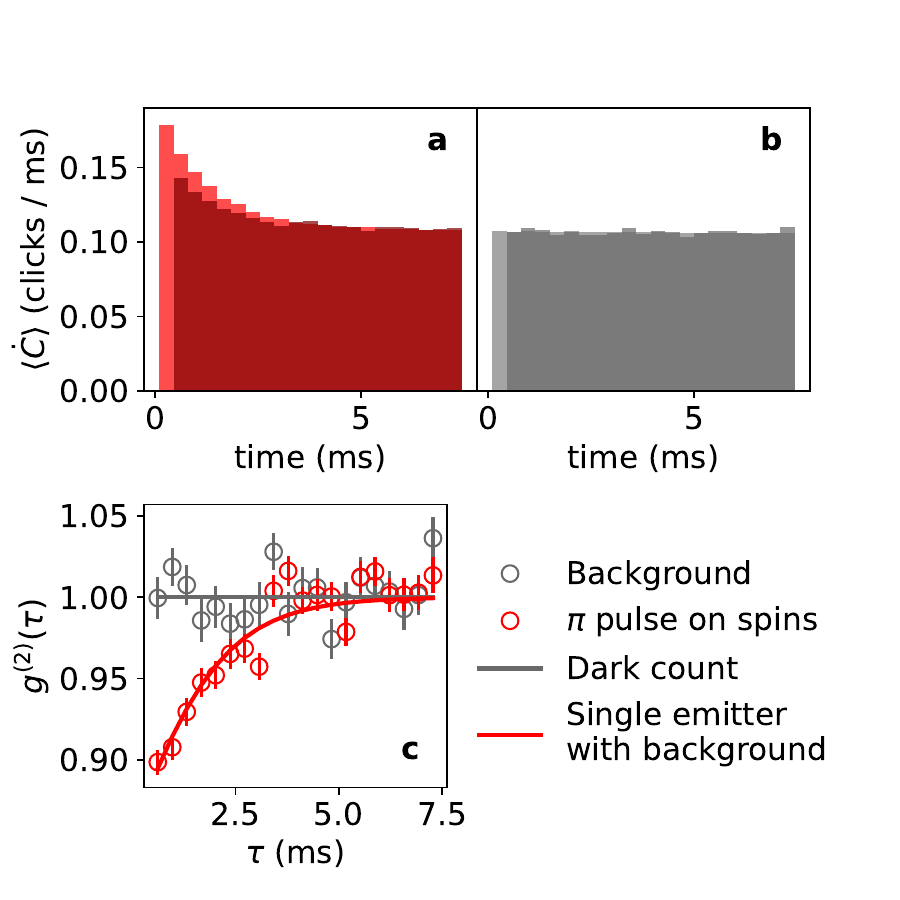}
\caption{
\textbf{Photon intensity auto-correlation function $g^{(2)}$ within one sequence.}
\textbf{(a)} Average count rate $\langle \Dot{C} \rangle$ as a function of delay time after a $\pi$ excitation pulse, for all recorded sequences (red) and for sequences with a first click detected before 0.45ms (dark red). The reduction of $\langle \Dot{C} \rangle$ in the second case indicates the anti-bunching of spin fluorescence photons. \textbf{(b)} Average count rate $\langle \Dot{C} \rangle$ as a function of delay time, for all recorded background traces (gray) and for traces with a first click detected before 0.45ms (dark gray). The unchanged $\langle \Dot{C} \rangle$ in the second case indicates a Poissonian background made of independent dark count events. \textbf{(c)} Extracted $g^{(2)}$ fucntion for dark counts (gray dots) and spin fluorescence signal (red dots) as a function of delay time $\tau$. Expected $g^{(2)}$ functions for the Poissonnian background (black solid line) and for an ideal single emitter in presence of the same background ($g^{(2)}_{se}$ - red solid line) fit well the experimental data. \textbf{(d)} Uncorrected $g^{(2)}(k)$ (blue columns) and corresponding $\pm$1-standard deviation error bars (red) as a function of inter-sequence offset $k$.
}
\label{fig_g2}
\end{figure}

We first provide in Extended Data Fig.~\ref{fig_g2}a a direct visualization of the anti-bunching found on a single-spin peak.  The count rate $\langle \dot{C} \rangle(t)$ is plotted as a function of time, first, averaged over all recorded sequences, and second, averaged over sequences with a count 1 in the first bin (conditioned curve). When measured on the background signal, the two curves are identical, whereas when measured on the single-spin peak ($s0$), the conditioned fluorescence rate is reduced at short times after the first count, compared to the average unconditioned one.

In order to quantify this anti-bunching, we then compute the intensity-intensity correlation functions inside a sequence,

\begin{equation}
    g^{(2)}(\tau=\tau_j)\frac{\langle n^{(i)}_0 n^{(i)}_j\rangle_i}{\langle n^{(i)}_0\rangle_i \langle n^{(i)}_j\rangle_i},
\end{equation}

\noindent as well as between two sequences separated by k excitation pulses,

\begin{equation}
    g^{(2)}(k)=\frac{\langle n^{(i)}_0 n^{(i+k)}_1+n^{(i)}_1 n^{(i+k)}_0\rangle_i/2}{\langle n^{(i)}_0\rangle_i \langle n^{(i+k)}_1\rangle_i},
\label{eq:g2_inter_pulse}
\end{equation}
where we keep only the first and second bin of the two sequences, symmetrize the function about k=0, and average over all pairs of sequences with same separation $k\in \mathbb{Z}$.

The intra-sequence $g^{(2)}(\tau)$ and inter-sequence $g^{(2)}(k)$ are shown in Extended Data Fig.~\ref{fig_g2}c and d. The latter is then corrected from the background counts (leading to Fig. 4c in the main text) as explained below.

\subsubsection{Background correction}


We now describe how we subtract from $g^{(2)}$ the dark count rate contribution, in order to obtain a background-corrected correlation function.

We assume that the clicks from the detector have two independent origins: emission $s_{j}$ from the spins, and Poisonnian background noise $d_{j}$ due to independent dark count events, such that $n_j=s_j+d_j$, $\langle n_j\rangle =  \langle s_j\rangle + \langle d_j \rangle $, and $\langle s_jd_j\rangle=\langle s_j\rangle\langle d_j\rangle$. In addition, we assume that the instruments during the measurement time are stable enough so that the dark count rate is time-invariant: $\langle d_j\rangle=\langle d\rangle$.

We thus define the background-corrected autocorrelation function
\begin{equation}
 g^{(2)}_{corr}(k)=\frac{\langle s^{(i)}_0 s^{(i+k)}_1+s^{(i)}_1 s^{(i+k)}_0\rangle_i/2}{\langle s^{(i)}_0\rangle_i \langle s^{(i+k)}_1\rangle_i}
\end{equation}
and express it explicitly as a function of the uncorrected $g^{(2)}(k)$ of Eq. \ref{eq:g2_inter_pulse} and of the measurement outcomes $A_j\equiv(\langle n_j^{(i)}\rangle_i - \langle d\rangle)/\langle d\rangle$:

\begin{equation}
    g^{(2)}_{corr}(k)=\frac{(1+A_0)(1+A_1)g^{(2)}(k)-A_0-A_1-1}{A_0A_1}.
\end{equation}


In addition, it is interesting to compare the measured $g^{(2)}(\tau)$ inside a sequence with the expected $g^{(2)}_{se}(\tau)$ that an ideal single emitter would give in presence of background noise. In this case, all terms $s_0^{(i)} s_{j}^{(i)}$ are 0 due to the single emitter character, and
\begin{equation}
    g^{(2)}_{se}(\tau=\tau_j) = \frac{\langle n_0^{(i)}\rangle_i\langle d\rangle+\langle n_{j}^{(i)}\rangle_i\langle d\rangle - \langle d\rangle^2}{\langle n_0^{(i)}\rangle_i\langle n_{j}^{(i)}\rangle_i}.
\end{equation}
This function is plotted as a red solid line in Extended Data Fig. \ref{fig_g2}(c) and shows a good match with the measured $g^{(2)}(\tau)$.


\subsection{Single-spin signal-to-noise ratio}

We derive in this section the theoretical single-spin signal-to-noise ratio of our fluorescence-detection protocol. This protocol involves identical sequences repeated every $t_r$ times during a total measurement time $t_m$. In each sequence, a $\pi$ pulse is applied at the beginning, and the number of counts is measured during a time $t_d$ following the pulse. The spin relaxation rate is $\Gamma_R$.

In our model, it is readily shown that the steady-state spin polarization at the beginning of each sequence is $S_{z0} = - \frac{1}{2} \tanh(\Gamma_R t_r/2)$, yielding an average number of counts per sequence $- 2 \eta S_{z0} (1 - \exp^{- \Gamma_R t_d})$, and an average total number of counts $C_{spin} = \eta(t_m/t_r) \tanh(\Gamma_R t_r/2) (1 - \exp^{- \Gamma_R t_d})$.

The noise has two contributions: one from the dark count fluctuations, whose variance is $\alpha t_d t_m / t_r$, and one from the partition noise of the detected photons, with variance $(1- \eta) C_{spin} $. Therefore, the width of the histogram with $\pi$ pulse is $\delta C_\pi = \sqrt{\alpha t_d t_m / t_r + (1-\eta) C_{spin}}$. The signal-to-noise ratio is defined as $SNR = C_{spin} / \delta C_\pi =C_{spin} / \sqrt{\alpha t_d t_m / t_r + (1-\eta) C_{spin} }$. 

For the parameters of our experiment ($\Gamma_R = 700 \mathrm{s}^{-1}$, $\alpha = 10^2 \mathrm{s}^{-1}$, $\eta = 0.12$), numerical optimization indicates a maximum SNR of $2.5$ is obtained for $t_d = 2\,\mathrm{ms}$ and $t_r = 3\,\mathrm{ms}$. In the experiment, we use a larger repetition time to minimize the effect of heating; for the parameters used ($t_d = 2\,\mathrm{ms}$ and $t_r = 7.5\,\mathrm{ms}$), the formula yields a SNR of $1.95$, in agreement with the measured value of $1.91$.

A simpler approximate scaling formula is obtained considering that the repetition time is $t_r \sim \Gamma_R^{-1}$, and that the spin polarization $S_{z0} \sim -1/2$. Taking moreover $t_d = t_r$, one obtains the scaling formula provided in the introduction for the $t_m=1\,\mathrm{s}$ integration time, $SNR \sim \eta \Gamma_R / \sqrt{\alpha + (1-\eta) \eta \Gamma_R}$.

\subsection{Non-radiative relaxation}
\begin{figure}[!tbh]
\includegraphics[width=0.3\textwidth]{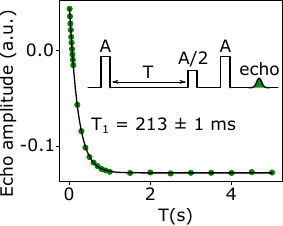}
\caption{
Spin-Lattice relaxation time measured with inversion-recovery sequence at 10 mK, with $B_0$ along the c axis, and $\omega_0/2\pi=7.853$\,GHz. Green dots are data, solid line is a fit yielding $T_1 = 0.213 \pm 0.001$\,s.
}
\label{fig_SpinLatticeT1}
\end{figure}

\begin{figure*}[!tbh]
\includegraphics[width=0.85\textwidth]{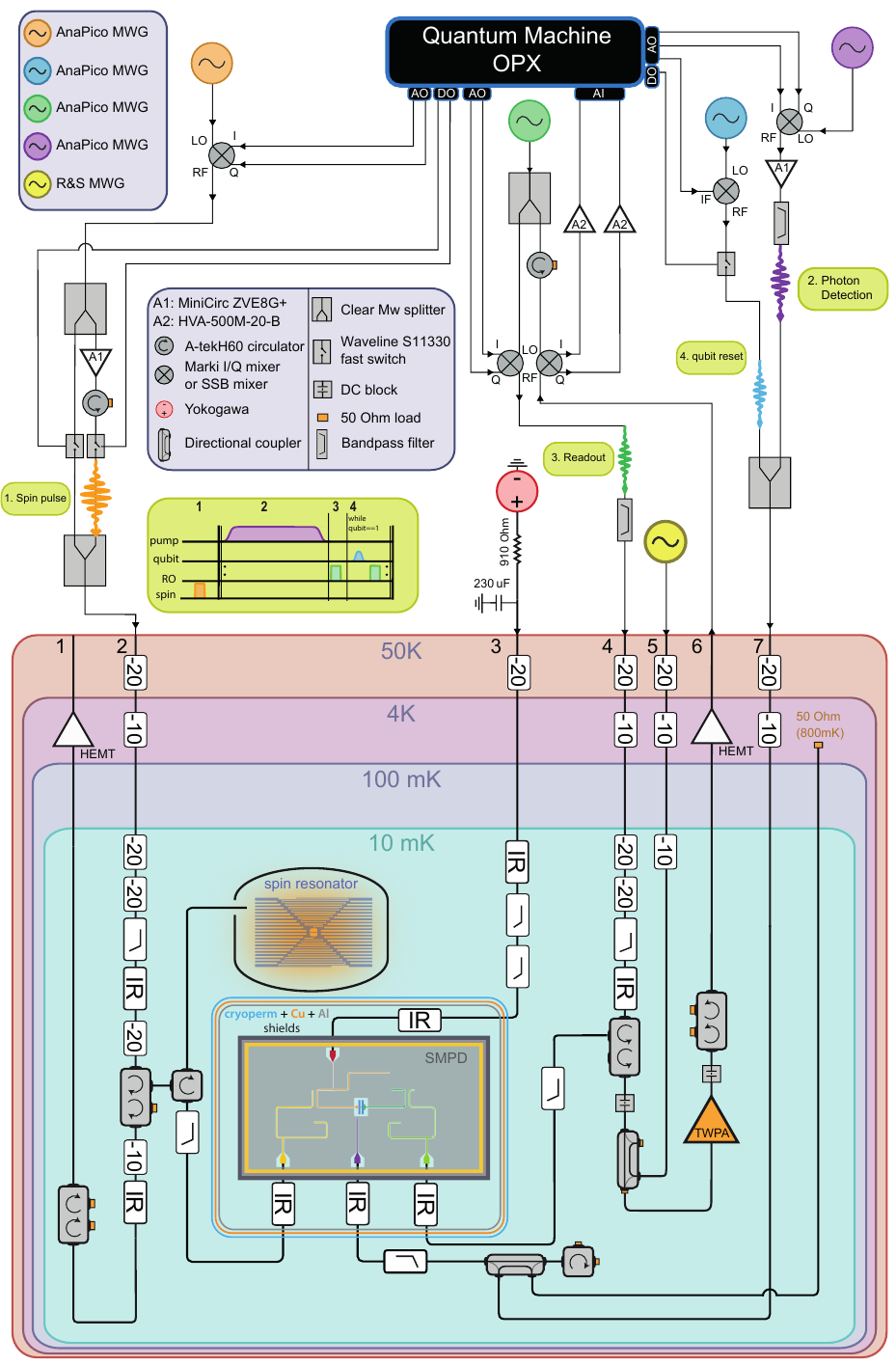}
\caption{
\textbf{Schematic of the setup.} Wiring and all the components used in this experiment at room temperature and cryogenic temperature are shown. 
}
\label{sub_fig1}
\end{figure*}

The spin-lattice relaxation time of $\mathrm{Er}^{3+}\mathrm{:CaWO}_4$ with $B_0$ oriented along the $c$ axis was measured using the traditional inductive detection, using a spin-echo-detected inversion-recovery sequence. A relaxation time $ T_1 = 210$\,ms was found at a frequency of $7.853$\,GHz~\cite{le_dantec_electron_2022} (see Fig. \ref{fig_SpinLatticeT1}). At $10$\,mK, the relaxation is dominated by the direct phonon process, with a non-radiative relaxation rate $\Gamma_{NR}$ scaling as $B_0^2 \omega_0^3$~\cite{larson_spin-lattice_1966}. Therefore, we estimate that in our conditions ($\omega_0/2\pi = 7.335$\,GHz, and $B_0$ applied along the $c$ axis), the non-radiative relaxation rate should be $\Gamma_{NR} \simeq 3.3\,\mathrm{s}^{-1}$.

\subsection{Efficiency}
We now discuss the value measured for the overall efficiency $\eta = 0.12$. Losses of counts can occur due to non-radiative spin relaxation, internal losses of the spin resonator, microwave losses between the spin device and the SMPD $\eta_{loss}$, and finite SMPD efficiency, so that $\eta = [\Gamma_R / (\Gamma_R + \Gamma_{NR})] [\kappa_c/\kappa] \eta_{loss} \eta_{SMPD}$. Given the measured $\eta_{SMPD}=0.32$, $\kappa_c/\kappa=0.57$, and $\Gamma_R / (\Gamma_R + \Gamma_{NR}) = 0.995$ we deduce $\eta_{loss} = 0.66$, which is a reasonable value for the microwave losses encountered upon propagation along a 50-cm-long coaxial cable, a circulator, and the filters at the SMPD input.

\subsection{Summary of different spins}
Apart from the spins discussed in the main text, we have also measured other single spin peaks found in the spectroscopy measurement. s0 and s6 are the labelled spins while s7 and s8 are not indicated in the spectrum in Fig. 2. Here we summarize their measured coherence time in the table below.
\begin{table}[!tbh]
\begin{tabular}{|c c c c|} 
\hline
Spin & $\mathrm{T_1(ms)} $ & $\mathrm{T_2^* (\mu s)}$  & $\mathrm{T_2^{echo}(ms)}$\\
\hline
s0 & 1.26 & 79 & 1.38\\
s6 & 1.42 & 170 & 2.47\\
s7 & 2.21   & 7.5 & 2.1\\
s8 & 1.36 & 315 & 1.53\\
\hline
\end{tabular}
\end{table}

\bibliographystyle{naturemag}
\bibliography{supplement}

\clearpage